# Cellular decision-making bias: the missing ingredient in cell functional diversity

**Bradly Alicea**[1]

[1]Orthogonal Research and Department of Animal Science, Michigan State University

**Abstract**

Cell functional diversity is a significant determinant on how biological processes unfold. Most accounts of diversity involve a search for sequence or expression differences. Perhaps there are more subtle mechanisms at work. Using the metaphor of information processing and decision-making might provide a clearer view of these subtleties. Understanding adaptive and transformative processes (such as cellular reprogramming) as a series of simple decisions allows us to use a technique called cellular signal detection theory (cellular SDT) to detect potential bias in mechanisms that favor one outcome over another. We can apply method of detecting cellular reprogramming bias to cellular reprogramming and other complex molecular processes. To demonstrate scope of this method, we will critically examine differences between cell phenotypes reprogrammed to muscle fiber and neuron phenotypes. In cases where the signature of phenotypic bias is cryptic, signatures of genomic bias (pre-existing and induced) may provide an alternative. The examination of these alternates will be explored using data from a series of fibroblast cell lines before cellular reprogramming (pre-existing) and differences between fractions of cellular RNA for individual genes after drug treatment (induced). In conclusion, the usefulness and limitations of this method and associated analogies will be discussed.

**Introduction**

Reprogramming cells from one phenotype to another is a well-established technique (see Vierbuchen and Wernig, 2012). Using a small number of factors, cells can be converted to a wide variety of types, including pluripotent cells (Park et.al, 2008; Takahashi et.al, 2007), neurons (Caiazzo et.al, 2011; Vierbuchen et.al, 2010; Pang et.al, 2011, Pfisterer et.al, 2011), muscle fibers (Davis, Weintraub, and Lassar, 1987), and cardiomyocytes (Qian, 2012). While significant characterization has been done in terms of efficiencies and candidate genes for induced function (Alicea et.al, 2013; Park and Daley, 2008), less explored is the differential capacity for a single cell line to convert to different phenotypes (e.g. neuron or muscle).

Are all forms of conversion equal, or are certain types of conversion easier to achieve? To better understand this question, these phenomena must be placed in the framework of more general biological mechanisms. In this paper, it will be argued that there is a clear bias towards some types of conversion over others. It will be further argued that this bias is variable across cell lines, and that this capacity is independent of priming by gene expression or the presence of precursor cells. This leads us to propose four hypotheses: the **phenotypic bias** (bias manifest in morphological indicators) hypothesis, the **pre-existing bias** (bias manifest in existing variation) hypothesis, the **induced bias** (bias manifest as a consequence of transformation) hypothesis, and the **extrinsic bias** (bias consistent with cell survivability under defined conditions).

Reprogramming bias is the tendency for some cell lines to favor a certain destination phenotype upon reprogramming (Figure 1). In this paper, we will compare two types of reprogramming: induced neuron (iNs) and induced skeletal muscle (iSMs). Across replicates, there is differential relative reprogramming efficiency between the two target phenotypes. We propose that there is a generic and inherent mechanism that contributes to a bias among certain input cells which results in a preference for some phenotypes over others (see Figure 1). While reprogramming bias is a formal measure that will be demonstrated on several small datasets, it is also intended to be a theoretical construct that may be linked to capacity for genomic change, phenotype-specific differential gene expression, and phenotype-specific regulatory patterns.

**Proof-of-Concept**

We will demonstrate the concept of reprogramming bias using a number of approaches. At its heart, the detection of reprogramming bias is done using a distribution-based assay that relies on a discrete version of signal detection theory (SDT – see MacMillan, 1999). An analysis of cell line survivability using an idealized model of environmental selection and the distribution-based assay (see Methods) will be compared with the distribution-based assay to reveal secondary effects. Taken together, these techniques make contributions to a theory of cellular decision-making.

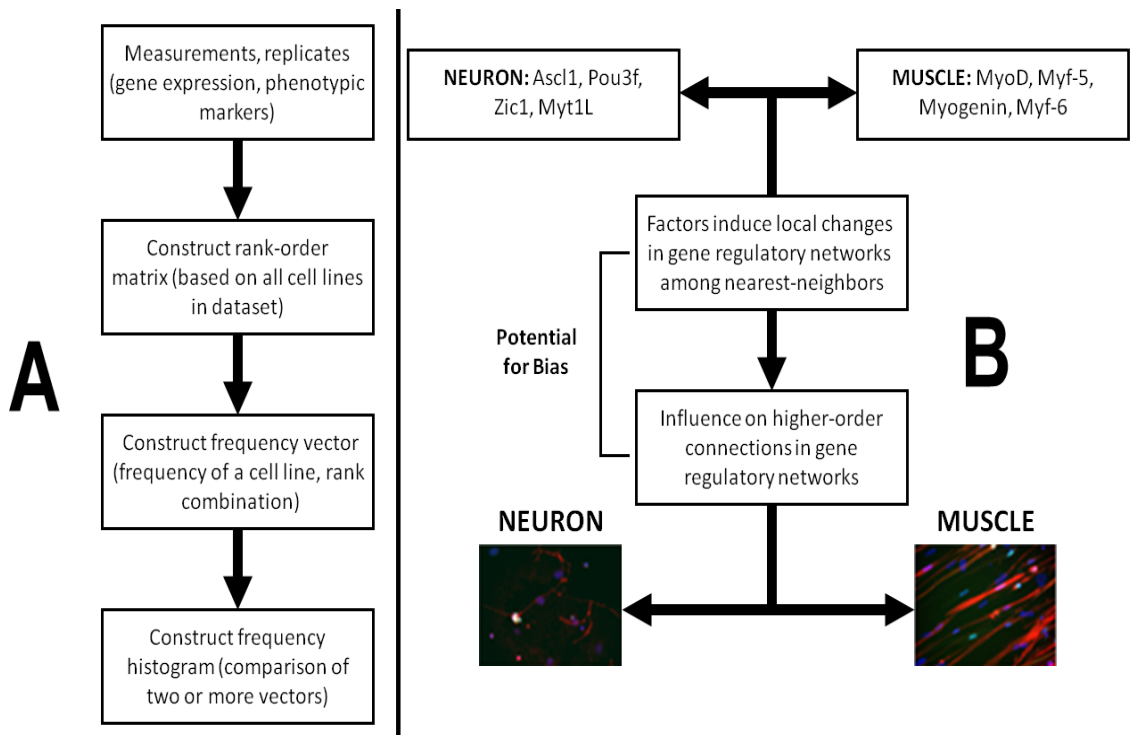

**Figure 1.** Meausrement and mechanisms of cellular reprogramming bias. **A:** Process diagram demonstrating how cell reprogramming bias is measured. **B:** Process diagram demonstrating the proposed cellular mechanisms behind reprogramming bias.

## Conceptual Overview of Cellular Bias

The candidate biological mechanism behind reprogramming bias builds in part on the idea of *developmental bias*. According to Arthur (2004), there are two roles for bias in developmental processes. In this case, bias acts to tune an existing genome in the service of environmental adaptation. The first role for bias is buffering, or the active suppression of phenotypic diversity. An additional role for bias involves the production of phenotypes that lead to novel behaviors using developmental selection. In both cases, bias can be viewed as a process which involves proximate selection on the inherent diversity of a cell (Smith et.al, 1985). During this process, the expression of a cell's genome is selectively constrained, resulting in the directional expression of this diversity.

Thus, reprogramming bias can be viewed as a type of developmental constraint (Arthur, 2001; Alberch, 1982). Psujek and Beer (2008) view bias via constraints as having both local and global components. Local bias, or genotype-dependent bias, results in different phenotypes produced from the same genotype. Global bias, by contrast, emerges from mutations in the regulatory structure of the genotype. In this sense, bias is not an active mechanism but a passive one that results from interactions with the environment.

Reprogramming bias also builds on the analogy of *decision-making bias*. To bring this into clearer perspective, the cellular reprogramming process can be modeled as a differentiation tree (Artyomov, Meissner, and Chakraborty, 2010) which unfolds according to a series of branching processes. This tree structure is analogous but not identical to developmental processes. During this branching process, a cell makes choices which are contingent upon environmental and regulatory cues. When these choices result in a non-random pattern, cells are said to be engaging in decision-making heuristics.

**Cellular decision-making: a framework**
Reprogramming bias also relies on the emerging theory of cellular decision-making. Cellular decision-making (see Balazsi, van Oudenaarden, and Collins, 2011) can be defined as cellular behaviors emerging from given certain environmental or regulatory contexts that result in the selection of some phenotypic states over others. Rather than propose that decision-making cells possess cognitive abilities, we will remind the reader as to the definition of the Latin root "cogno", which means "to know". As is the case with the cognitive theory of nervous systems, the implication of "to know" translates into performing operations on (or processing) environmental information (Bechtel and Herschbach, 2009). In the case of cellular SDT, discrete changes in regulatory response lead to differences in performance, which in turn leads to a certain degree of measurable bias.

The basic cellular decision-making model involves deviation from random expectation. As is the case with developmental bias, this implies a selection mechanism. However, in cellular decision-making, the bias may involve basic mechanisms used as information-processing heuristics. These need not require sophisticated sources of memory or deterministic planning on the part of the cell. One example of this is stochastic switching in bacterial cells (Beaumont, Gallie, Kost, Ferguson, and Rainey, 2009) which is similar to cognitive forms of bet hedging. While in both cases this might result from a fluctuating environment (Samoilov, Price, and Arkin, 2006), the cellular version involves a simple mechanism tied to environmentally-triggered, gene expression regulation organized in a combinatorial fashion (Acar, Mattetal, and van Oudenaarden, 2008; Schultz, Lu, Stavropoulos, Onuchic, and Ben-Jacob, 2013; Bijlsma and Groisman, 2003).

**SDT and the detection of discriminability**

While simple examples of binary decision-making can be found among regulatory motifs, the aggregate outcomes of gene regulatory network activity can also be modeled and analyzed using a decision-making framework. The analysis of reprogramming bias and its consequences is inspired by traditional SDT theory used in psychophysics (MacMillan, 1999). This approach (cellular SDT) utilizes a modified version of traditional SDT to better capture the nature of cellular information processing and decision-making during cellular reprogramming and other forms of transformation. SDT describes the ability of a receiver to discriminate a signal embedded in a background of noise, and models signal and noise as potentially separable Gaussian distributions (Hutchinson, 1981). The degree of overlap between these two distributions (signal and noise) determines the discriminability of the decision-making mechanism. The greater the separability between signal and noise, the more discriminative power a specific decision-making mechanism possesses.

*Bias as discriminability in cellular SDT.* In cognitive decision-making, this discriminative power is improved with more environmental information. As in the case of cognitive decision-making, environmental information is critical to making preferential decisions. Cellular SDT is based on a single cell line evaluative technique (Alicea, 2013) that involves rank-ordering cell lines in a dataset with regard to their performance, and then calculating the frequency of all rank positions across replicates (see Supplemental Figure 1). In cellular SDT, the basic assumptions of traditional SDT and this evaluative technique are combined to find potential discriminability (e.g. bias) between different measures, functional categories of gene, and types of reprogramming.

This theory is also useful for understanding why a given cell type will preferentially convert to one phenotype over another. There are notable differences from a conventional SDT model (Figure 2). In cellular SDT, the comparison is not between distributions of signal and noise, but between distributions for two alternate signals (e.g. phenotypic conversion to neuron and muscle). The distinction (e.g. bias) is assessed by the degree of separability between the two distributions. This corresponds with the discriminability parameter (d') in traditional SDT theory that characterizes the ability to discriminate signal from noise. Cellular SDT can thus be distinguished in two ways: there is no calculation of receiver operating characteristics (ROC) nor is the concept of bias based on correct rejections and false alarms. Nevertheless, the rank-order criterions provides us an alternate window into decision-making: constancy and variability across replicates.

Perhaps a more important difference between the two models is that in cellular SDT, the distributions are coarse-grained frequency histograms that can derive from either rank-order (e.g. ordinal) measurements or binned (e.g. discretized) continuous measurements. As a binomial measure, the overlap between these distributions (e.g. histograms) can be directly contrasted with the degree of reprogramming bias exhibited between individual cell lines. In short, the less overlap there is between two frequency distributions, the greater the bias. And while there is no distribution that explicitly represents noise, high degrees of overlap or the lack of support in a generated distribution can be used to approximate high degrees of noise.

*Burstiness and Bias*. From changes in phenotype (Tosh and Horb, 2004) to cascades related to gene expression and epigenetics (Richard et.al, 2011), cellular reprogramming can be defined as a series of large-scale effects arising from small-scale perturbations. One unresolved issue regarding the basic biology of direct reprogramming involves statistical signatures of variation that might help us better understand the non-uniformity of this process (Hanna, Saha, and Jaenisch, 2010).

To demonstrate this, it is shown that reprogramming events (defined using a time-aggregate measure of efficiency) can be characterized as a Poisson process. Conceptually, the bursty nature of cellular reprogramming can be understood by indirectly observing these events as variability in reprogramming efficiency. The proportion of cells in a population that convert to a new phenotype (e.g. reprogramming efficiency) has been observed to vary from 0.001% to 29% for induced pluripotent stem cell (iPSC) reprogramming (Artyomov, Meissner, and Chakraborty, 2010).

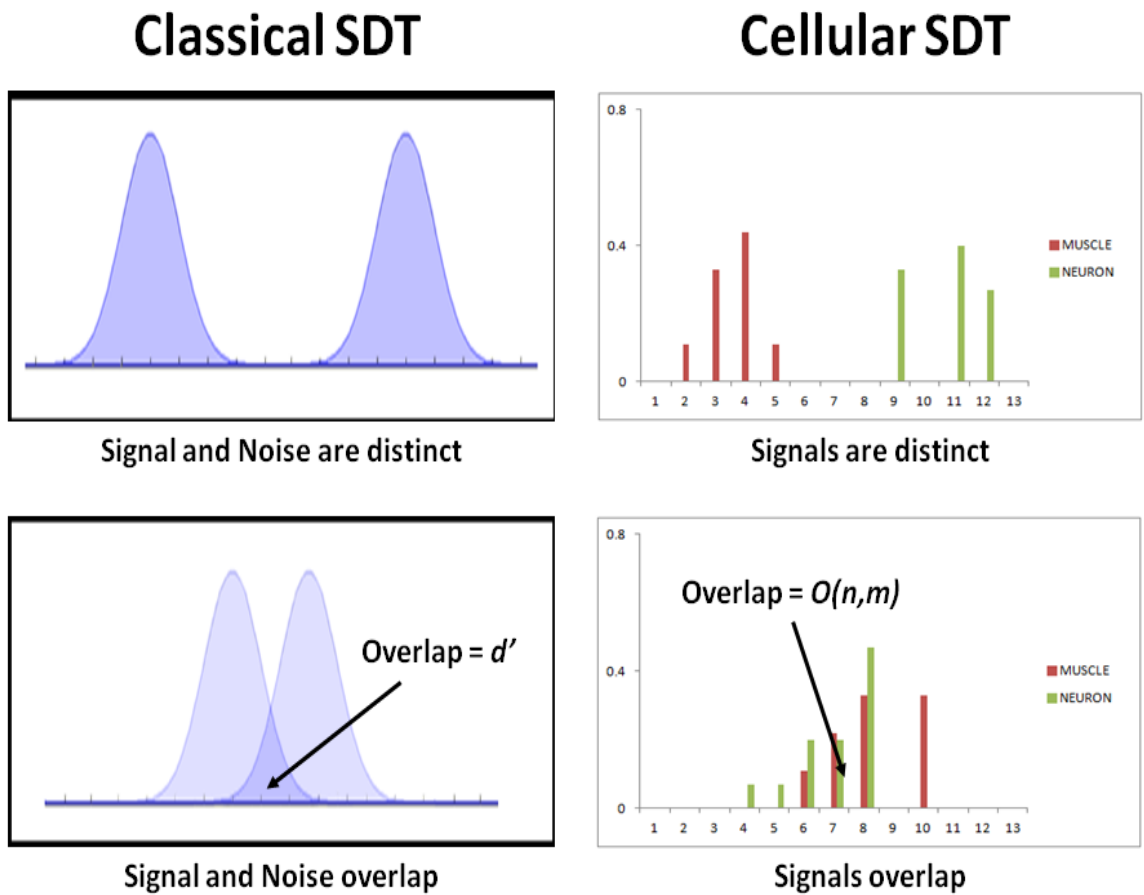

**Figure 2. LEFT:** Classical SDT (human decision-making). Idealized Gaussians used for demonstration purposes only. **RIGHT:** Cellular SDT (cellular decision-making). Possible distributions include Gaussian, Exponential, and Multimodal.

By contrast, bursty behavior of human (Barabasi, 2005) and physical (Karsai, Kaski, Barabasi, and Kertesz, 2012) systems is generally understood as a series of short intervals of high activity interspersed among longer periods of low activity. In the context of direct reprogramming, many-fold differences in reprogramming efficiency between cell lines (see Alicea et.al, 2013) can be defined as bursty behavior with no clear contribution of genetic background or conventional sources of biological variation. Cellular SDT may allow us to get at the sources of this variation, and assign it (at least in part) to the aggregate information-processing functions of cells.

### Reprogramming Bias: a theoretical example

In previous experiments (Alicea et.al, 2013), it has been shown that some cell lines demonstrate a clear preference over other lines for reprogramming to a single target phenotype. This preference has been demonstrated to be a property of fibroblasts populations with neither progenitor cell influence nor cells which are primed for gene expression reflecting the destination phenotype. This phenomenon can be shown in a contingency table (Table 1), in which all possible outcomes are featured for a single cell line exposed to both the muscle and neuronal reprogramming factors.

In this hypothetical set of experiments, a cell line that exhibits high reprogramming efficiency for both iSM and iN phenotypes is said to exhibit a generalized plasticity mechanism. This may involve changes in cell cycle timing or other changes to generalized cellular mechanisms (Egli, Birkhoff, and Eggan, 2008; Cox and Rizzinio, 2010). When a cellular population exhibits low reprogramming efficiency for both iSM and iN phenotypes (the diagonal cells in Table 4.4), this suggests active suppression of phenotypic change related to developmental buffering mechanisms (Chipev and Simon, 2002; Blasi et.al, 2011). However, when a cell population exhibits a high reprogramming efficiency for one phenotype (e.g. iN) but not another (e.g. iSM), it is proposed that these cells exhibit reprogramming bias.

**Table 1.** All possible outcomes for experiments reprogramming cells to induced neuron (iN) and induced skeletal muscle fiber (iSM) phenotypes. (+) is equivalent to above-average reprogramming efficiency, while (-) is equivalent to below-average reprogramming efficiency.

|  | **iN indicator (+)** | **iN indicator (-)** |
|---|---|---|
| **iSM indicator (+)** | Generalized plasticity | iSM Bias |
| **iSM indicator (-)** | iN Bias | Active suppression/buffering |

The driving mechanism behind reprogramming bias is a preference towards changes specific to the genetic regulatory network of a given type. When a cell is exposed to the reprogramming factors, there is a reordering of the cell biochemistry that allows for major phenotypic changes to occur. Reprogramming bias is simply a functional directionality in these changes. For example, iSM bias will involve a preference for changes specific to the skeletal muscle regulatory network (Bismuth and Relaix, 2010). By contrast, iN bias will involve changes to specific types (e.g. glutamitergic) of neuronal regulatory network (Hobert, 2008).

In terms of cellular populations, biased cells are non-progenitor cells that favor one specific phenotype (e.g. iSM) over another (e.g. iN). If enough individual cells in the population meet this criterion, the population can be said to exhibit strong bias. Previous studies suggest that determination of this bias may be highly centralized, as Nanog fluctuations may control this bias in iPS and other stem-like cells (Kalmar et.al, 2009). In terms of a regulatory-specific process, the introduction of cell type-specific transcription factors will trigger an immediate response in a core set of genes, or a first-order genetic regulatory network.

## Results

To understand the context and limitations of cellular SDT and the phenomenon of cellular decision-making, results from five datasets will be consulted (see Methods). An analysis that demonstrates the bursty, non-uniform nature of reprogramming will set up a demonstration of both phenotypic and genomic bias.

### Distribution of Events in Reprogramming

To fully appreciate why reprogramming bias represents fundamental differences in reprogramming performance between biological contexts, we must first understand how reprogramming events are distributed in time. As an example of biologically bursty behavior (and shown empirically in Alicea et.al, 2013), it was demonstrated that several-fold differences exist for reprogramming efficiency between cell lines of different origins. Further analysis focusing on this finding reveals that these differences can be modeled using an exponential distribution. This exponential model corresponds to a Poisson arrival process (Consul, 1989; (Barbour, Holst, and Janson, 1992), which is a more general description of a stochastic process unfolding over time (see Figure 3). When these data are resampled and tested for significance (using a Poisson exact test) within individual cell lines, the results reveal that only some cell lines exhibit an exponential distribution of reprogramming efficiencies across replicates (Supplemental Figure 2).

### Reprogramming Bias in the Phenotype and Genotype

The theory of reprogramming bias can be applied to measurements of both the phenotype and genotype. Difference between phenotypic and genomic measurement is the sensitivity (trade-off) of measurement: phenotype is stronger signal but less precise, genotype is weaker signal but potentially more precise with respect to potential biasing mechanisms.

## Phenotypic Bias

Phenotypic reprogramming bias is measured for both muscle and neuron destination phenotypes for 13 mouse cell lines. While these lines are representative of 7 tissue types, they are homogeneous with respect to genotype. Yet as shown in Supplemental Figure 3 and Figure 4, there is significant variation in the degree of bias exhibited amongst these cell lines.

In the case of phenotypic measurements, we can exhibit bias as a function of differential but collective genomic regulation. By the time phenotypic indicators give us a clear signal of reprogramming efficiency, a large number of genomic events that contribute to differential reprogramming have already occurred. Therefore, while we should expect a larger effect using phenotypic measurements, we may also need to sacrifice our understanding of the mechanistic processes behind the observed bias.

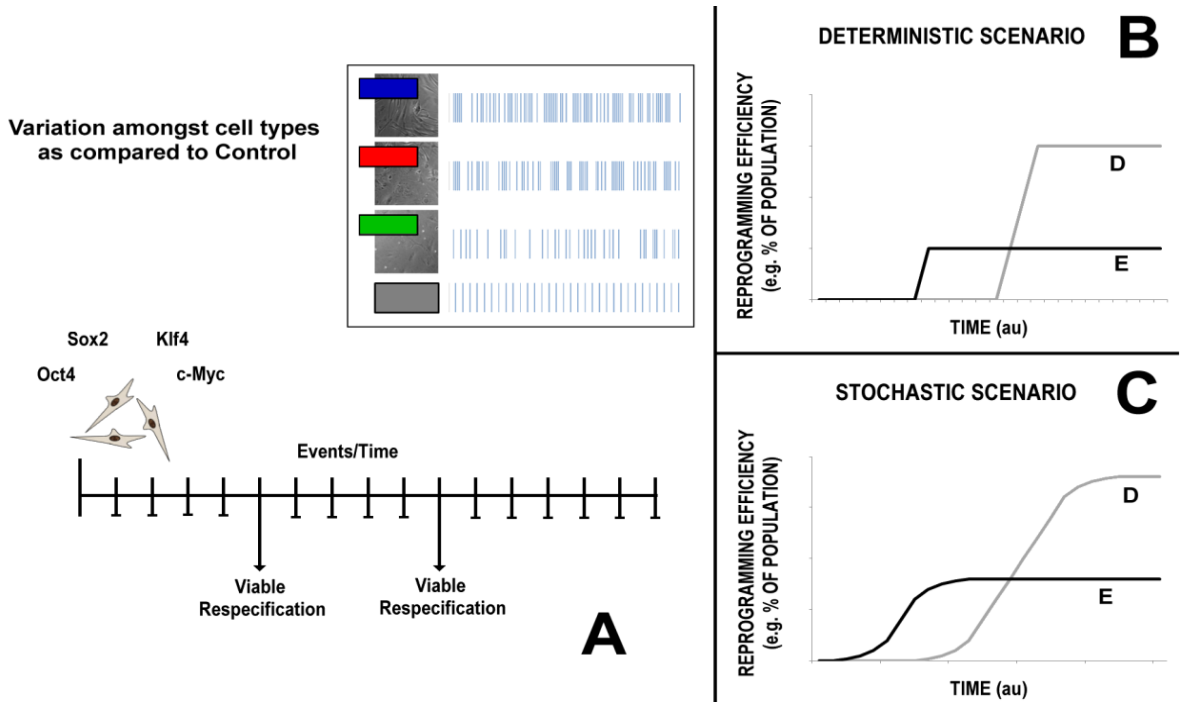

**Figure 3.** Current thinking about source of variation and the reprogramming process. A) Figure 1 taken in part from Sridharan and Plath (2008), and describes the role of variation across biology and over time; B) and C) are the deterministic and stochastic scenarios, respectively, taken from Hanna et al. (2009), Figure 1. Legend for schematic functions in B) and C): D – democratic, E – elite.

**Genomic Bias**

Genomic signatures of reprogramming bias can also be determined using cellular SDT. Bias may exist between the same gene in cells of two different phenotypes, or as a pattern across many genes. Unlike in the case of phenotypic bias, genomic bias can reveal the underlying mechanisms of reprogramming bias. However, particularly in the case of single-gene signatures of bias, the effect may be very weak. In cases where the biasing mechanism is based on a particular pattern of expression, bias can be used as an alternative way of revealing differential gene expression. In a related application, the bias measurement may be used as a way to distinguish between two different fractions of cellular RNA or as a way of better understanding the expression of alternative transcripts.

**Genomic Bias Test #1: pre-existing bias**

One alternative hypothesis involves asking whether or not the bias revealed in post-reprogramming phenotypes is simply an artifact of existing differences in gene expression within or between cell lines prior to reprogramming. If low overlap exists between two genes or cell lines, then this is evidence of pre-existing bias. Figure 5 shows an example of how this is done using four fibroblast cell lines from two tissue types of origin (tail tip and testes) assayed for gene expression patterns prior to reprogramming.

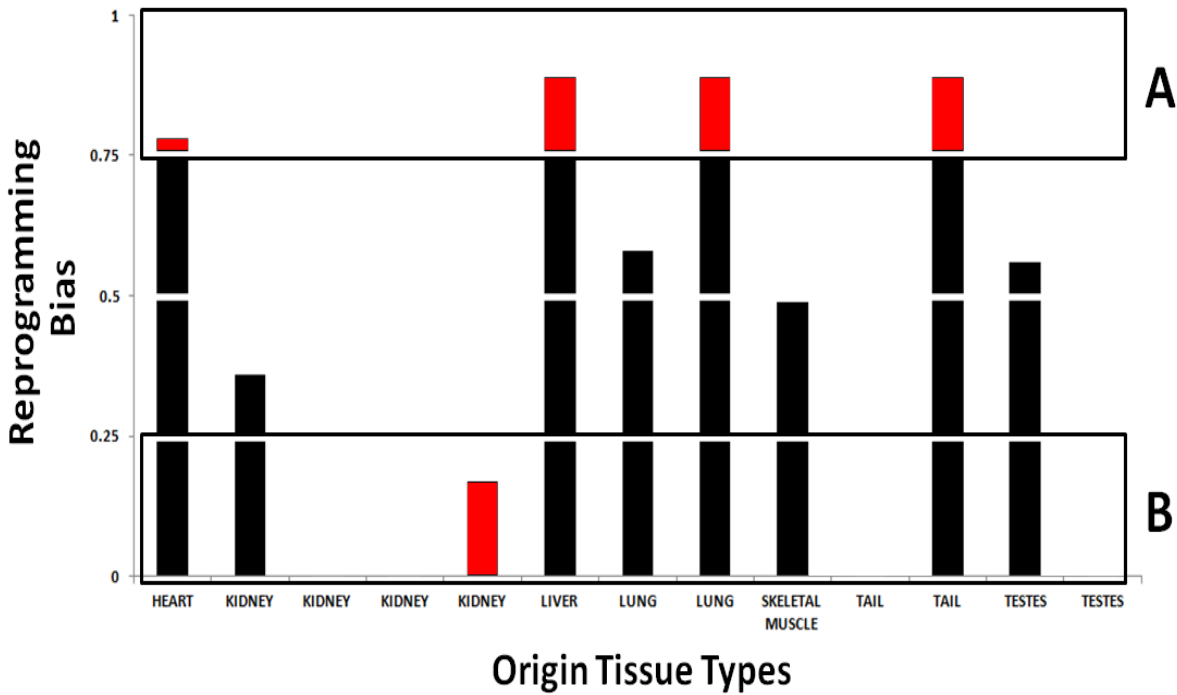

**Figure 4.** A bar graph showing the relative reprogramming bias for 13 mouse cell lines originating from different sources tissues; **A:** region of high bias (top quartile, total of 8 cell lines), **B:** region of low bias (lowest quartile, total of 1 cell line).

In this case, a very high degree of overlap exists between both cell lines of the same tissue type (insets) and cell lines from different tissue types (averaged together, main graph). The results in Figure 5 suggests that for the most part there is no consistent pattern of upregulation across all genes assayed (e.g. a greater frequency of high rank-order position), although this is nominally higher for one testes and one tail tip line. The error bars (standard error) in the main graph reveal that much of the variability among cell lines occurs in the first-place position of the rank-order. This may be due to the role of gene expression noise. However, this finding coupled with a lack of reprogramming bias may suggest that noise does not contribute to pre-existing reprogramming bias. Similar patterns have been confirmed using genes from high-throughput gene expression datasets annotated for neuron and muscle functions.

**Genomic Bias Test #2: induced bias**
Another alternative hypothesis involves the presence of genomic bias once a phenotype has been induced due to reprogramming or other treatment. In this case, measurements of transcription-associated RNA and translation-associated RNA can be examined for bias after various drug treatments. Figure 6 shows the potential for reprogramming bias for single genes among different fractions of cellular RNA.

In all three representative genes (COL1A, Fibronectin, UTF), there is significant overlap for transcriptional- and translational-associated RNA. However, gene expression for COL1A tends to be high-ranked, while gene expression for UTF tends to be low-ranked. This is true for both transcriptional- and translational-associated RNA. In the case of both COL1A and UTF, the distributions are spread out in a manner that suggests intermittent changes in gene expression across contexts. While these results have not been replicated using a high-throughput dataset, there are interesting local patterns which suggest intermittent changes in gene expression due to bias between fibroblast and pluripotent cell types.

**Survivability Analysis**

To answer the question of whether or not the ability of a cell to survive has any bearing on reprogramming bias, a separate dataset can be used to assess the survivability of a given cell line. These cell lines (evaluated as non-converted fibroblasts) are rank-ordered and a distribution is built as in the case of neuron and muscle data in Supplemental Figure 2. The survivability histograms can be compared with neuron and muscle data (Supplemental Figure 4), and the degree of overlap with these distributions can indicate whether survivability corresponds with conversion to a muscle phenotype, a neuron phenotype, or both.

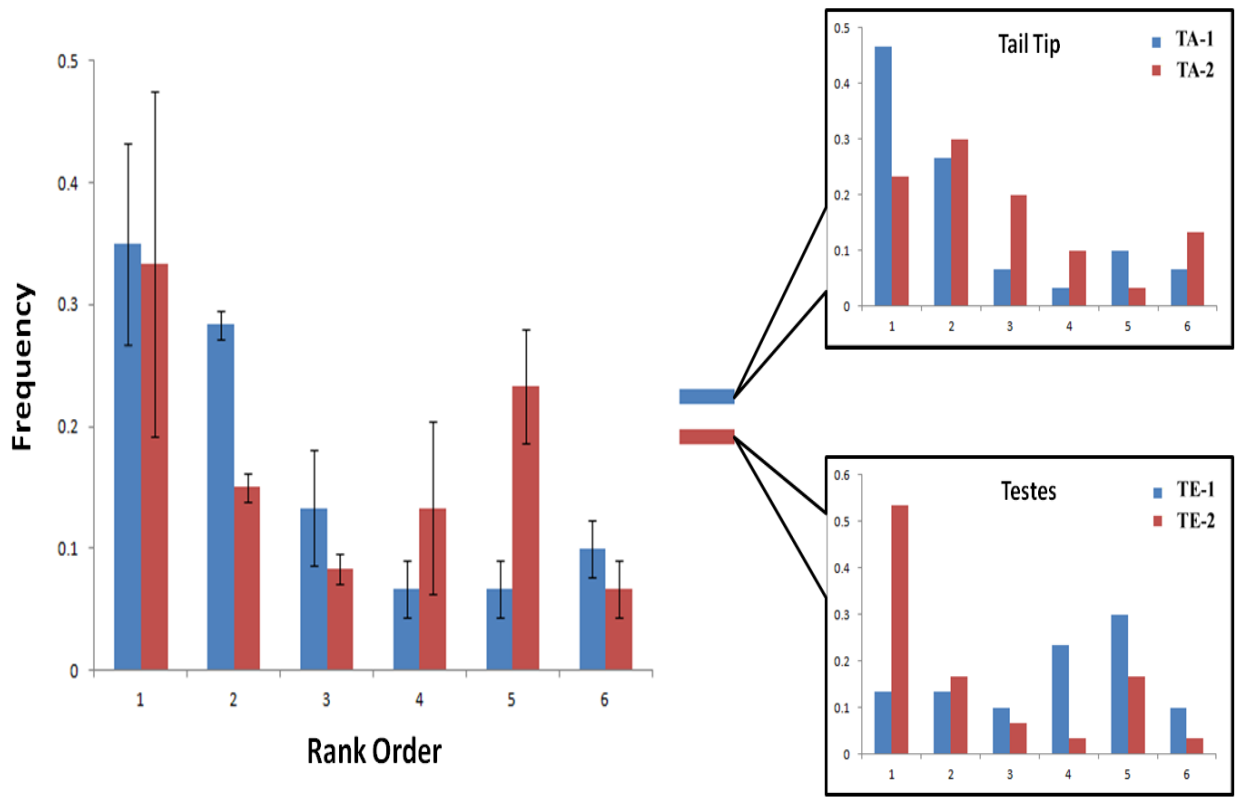

**Figure 5.** Examples of pre-existing bias detection, using four cell lines from two tissues of origin.

## Discussion

This demonstration can be interpreted in terms of premises related to what we have learned so far from looking at the role of variation in cellular reprogramming. In this work, there exists a many-fold difference in reprogramming efficiency between cells of the same genetic background. This cannot be explained by prior neuronal or muscle gene expression, as differential gene expression evaluated using a candidate gene approach yields no corresponding differences. This also cannot be explained by the presence of a progenitor cell population, as populations that exhibit the aforementioned effect are not positive for candidate markers.

**Collective decision-making**

This technique is explicitly geared towards populations and collective cellular processes. Cellular SDT is a way to extract collective decision-making from large populations of single cells and genomes. While many cellular decision-making studies focus on assumptions best suited for mechanisms in single cells, cellular SDT allows for a range of cellular mechanisms to be evaluated within the cellular decision-making framework. In addition, the detection of bias may only be one of many possible signals detectable using this framework. A wider range of applications might better determine its true degree of generalizability.

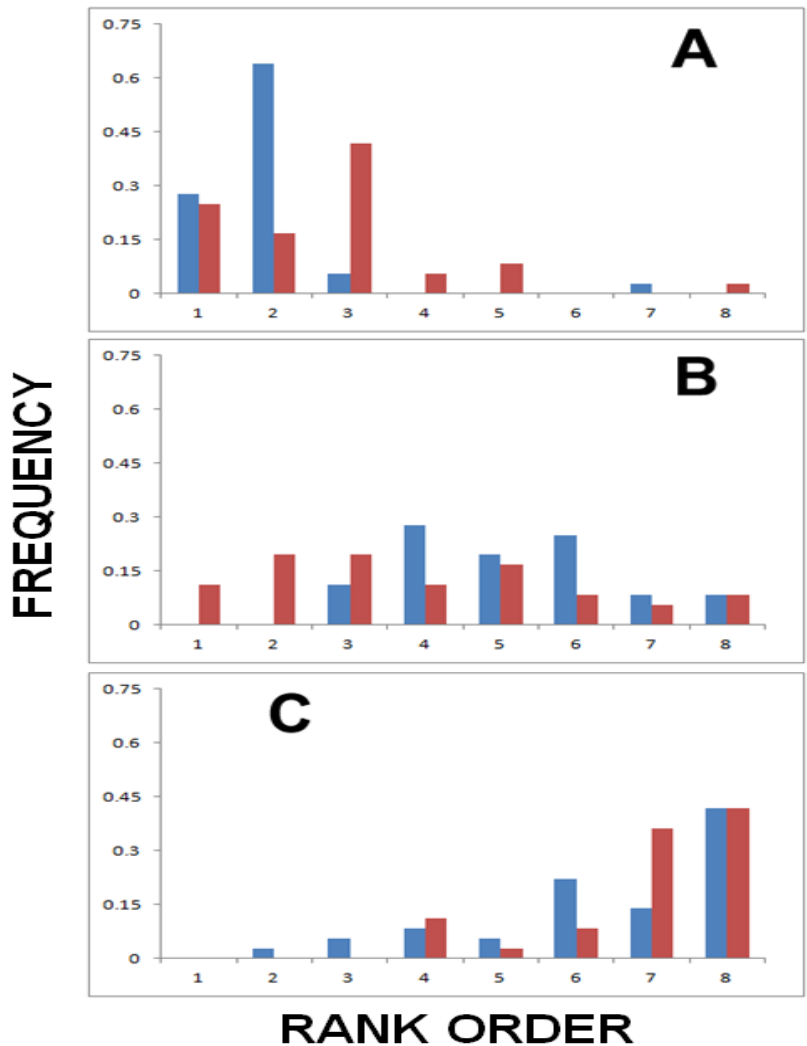

**Figure 6.** Examples of induced bias detection, comparing two fractions of RNA across genes with different patterns of baseline expression. **A:** COL1A (fibroblast-specific), **B:** Fibronectin (fibroblast-specific), **C:** UTF (non-specific). Transcription-associated RNA (red), translation-associated RNA (blue).

**Limitations and Broader Considerations**

While the bias measurement captures the basic decision-making outcomes rather easily, there are drawbacks to this technique. Since decision-making across a population of cells and across replicates is an aggregate, stochastic process, there is significant variability. In cases where the sample size (e.g. number of rank-order items) is relatively small, measurements of bias can be hard to estimate. This also holds true for samples with a weak experimental effect. However, in both cases, a distinction can be made between "tight" histograms (e.g. instances where almost every rank is in the same position) and "spread out" histograms (e.g. instances that vary across the rank order). Even in cases where samples overlap significantly, making the distinction between these two types of distribution can improve the resolution of this technique.

Figure 3 (panels B and C) demonstrates one of the advantages of using the cellular SDT approach to better understand cellular reprogramming. In this case, what has been referred to in this paper as reprogramming bias might rely upon a set of stochastic mechanisms (e.g. the stochastic scenario). In this case, not only are there variable latencies in the reprogramming process (time from infection to mature phenotype), but multiple transformational responses for the same genetic background as well. This might explain the variable bias in phenotypic potential given both the same genetic background and relative lack of bias in the genotype.

**Table 2.** Bias measurements for 13 fibroblast cell lines (mouse tissue of origin): survival rate vs. muscle reprogramming efficiency vs. neuron reprogramming efficiency.

|  | Survival-Muscle | Survival-Neuron | Difference | Muscle-Neuron |
|---|---|---|---|---|
| Heart-1 | 1.00 | 0.93 | 0.07 | 0.78 |
| Kidney-1 | 1.00 | 1.00 | 0.00 | 0.36 |
| Kidney-2 | 0.78 | 0.83 | 0.05 | 1.00 |
| Kidney-3 | 0.78 | 0.73 | 0.05 | 1.00 |
| Kidney-4 | 1.00 | 1.00 | 0.00 | 0.17 |
| Liver-1 | 1.00 | 0.86 | 0.14 | 0.89 |
| Lung-1 | 0.67 | 0.76 | 0.09 | 0.58 |
| Lung-2 | 1.00 | 0.80 | 0.20 | 0.89 |
| Skeletal Muscle-1 | 0.83 | 1.00 | 0.17 | 0.49 |
| Tail Tip-1 | 0.83 | 0.47 | 0.36 | 1.00 |
| Tail Tip-2 | 1.00 | 0.83 | 0.17 | 0.89 |
| Testes-1 | 1.00 | 0.86 | 0.14 | 0.56 |
| Testes-2 | 0.72 | 0.87 | 0.15 | 1.00 |

While quite different from conventional SDT, cellular SDT provides an opportunity to characterize phenomena that often go unmeasured in the analysis of gene expression and phenotypic data. For one, cellular SDT offers an alternative to the "group of genes go up, group of genes go down" approach of conventional differential gene expression analysis. In this case, differences in bias are hypothesized to be information processing differences, while differences not due to information processing result in overlap. While this provides us with a sparse, multimodal signal, local regularization and comparisons of measurement between data types (e.g. phenotypic and pre-existing bias) can help to resolve available information.

As a discrete variant of SDT, reprogramming bias can be used to capture decision-making related to cellular-level transformative processes. This may include cellular differentiation in development and evolution, cancer-related cell transformations, and the genomic underpinnings of cryptic phenotypic changes. Of particular interest is the operation of these processes across diverse cell populations (Altschuler and Wu, 2010). Patterns of information processing (and hence simple decision-making) within a specific context can be evaluated using the rank-order frequency method, and compared across contexts using cellular SDT. While the cellular SDT technique relies on accurate methods of quantification, a range of such methods can be easily adapted to a cellular SDT analysis. In terms of broader relevance, cellular SDT may allow us to understand the basic biology of reprogramming and induced phenotypic change for applications to regenerative applications and the study of cancer phenotypes.

## Acknowledgements

Thanks go to colleagues at the Cellular Reprogramming Laboratory Journal Club and the Regenerative Medicine Research Group (particularly Dr. Steven Suhr) for data and input. Additional thanks go to the BEACON Center at Michigan State, as this idea was also presented to the Genomes and Evolution group. Some datasets were collected as part a NINDS grant (#1RZ1NS076959-01A1). Dataset acknowledgements go to the GenBank and EMBL databases, which served as the source for the high-throughput datasets.

## Methods

**Datasets**

There were five datasets used in this paper. The first (to measure phenotypic reprogramming bias) involves mouse fibroblast cell lines after reprogramming to induced neuron and induced muscle. The second dataset involves qPCR measurements of mouse fibroblast cell lines before reprogramming. The third dataset involves human fibroblast cell lines, and consists of qPCR measurements of differences between translational-associated mRNA and transcriptional-related qPCR. The fourth dataset involves mouse fibroblast cell lines that are counted for survival after several days of exposure to defined conditions. The fifth dataset consists of microarray assays for multiple fibroblast and pluripotent cell lines. Futher details about these datasets can be found in Alicea (2013).

**Computer Code**

MATLAB code for rank-order analysis, histogram construction, and overlap/bias calculations can be found at the Github repository: https://github.com/balicea/cellular-SDT.

**Reprogramming Efficiency**

Reprogramming efficiency was calculated using the method of Alicea et.al (2013). Briefly, DAPI (cell body) and immunoflorescence markers were imaged from wells of iN and iSM cell lines. Reprogramming efficiency was the proportion of cell-type marker positive (red channel) to infection-marker positive (YFP channel) cells. All images were normalized by the total number of cells (DAPI channel).

**Bias Measurement**

Bias was calculated by rank-ordering all experimental replicates and then calculating a frequency matrix for all cell line, rank order ($j,k$) combinations. For each $i^{th}$ element of $j$, the overlap for destination cell types such as neuron ($n$) and muscle ($m$) can be directly compared using a histogram. This can be described mathematically as

$$O(N,M) = \Sigma \, MAX(N_i, M_i) - \|N_i - M_i\| \qquad [1]$$

The reprogramming bias for a specific genomic or phenotypic signal can be measured by calculating the inverse of this index for all non-zero values, which can be defined as

$$BIAS = \begin{cases} 1 - O(N,M) & \text{if } O(N,M) > 0 \\ \quad\quad\text{else } 0 \end{cases} \qquad [2]$$

**Survival Measurement**

The survival measurement was derived from a cell lines put under selective (e.g. death) conditions (conditioned media which challenges cell division and growth) for 6d. Survival was characterized by counting the difference between the number of cells (derived using cell counts from random samples) in a given culture at 0d and 6d.

**Reprogramming Regimens**

The creation of iN and iSM cells involves the regimens introduced in Alicea et.al (2013). Briefly, the iN cell lines are created using a retroviral vector containing four transcription factors (Ascl1, Pou3f, Zic1, and Myt1L). The iSM cell lines are created using a retroviral vector also containing four transcription factors (MyoD, Myf-5, Myogenin, Myf-6).

**Transcriptional-associated RNA**

Transcriptional-associated RNA is isolated using Trizol method. The resulting RNA is cleaned using a Qiagen kit. All mRNA for transcriptional-associated RNA (as well as translational-associated RNA) were quantified using Nanodrop spectroscopy and qRT-PCR.

**Translational-associated RNA**

The full method for isolating translatome-associated RNA was done using the protocol available in Alicea, Suhr, and Cibelli (in preparation). Briefly, RNA was obtained via homogenization using polyribosome extraction buffer (PEB). The polysome is isolated from lysate, and RNA is precipitated using sodium acetate. Magnetic-assisted cell sorting (MACS) can be used to provide a signal more specific to translational processes. The resulting RNA is cleaned using a Qiagen kit.

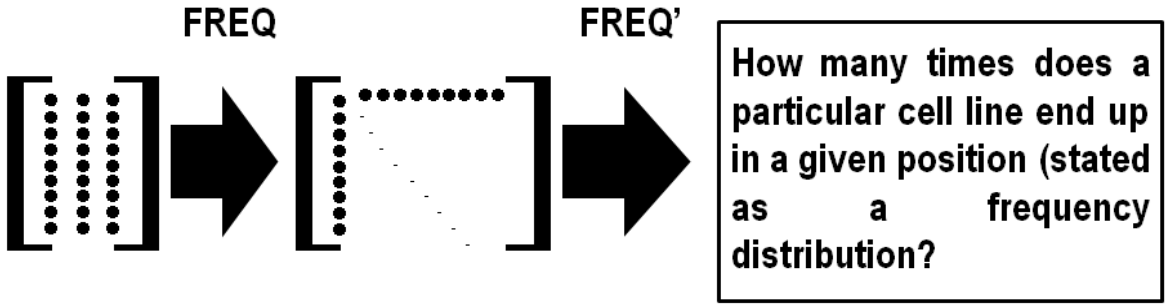

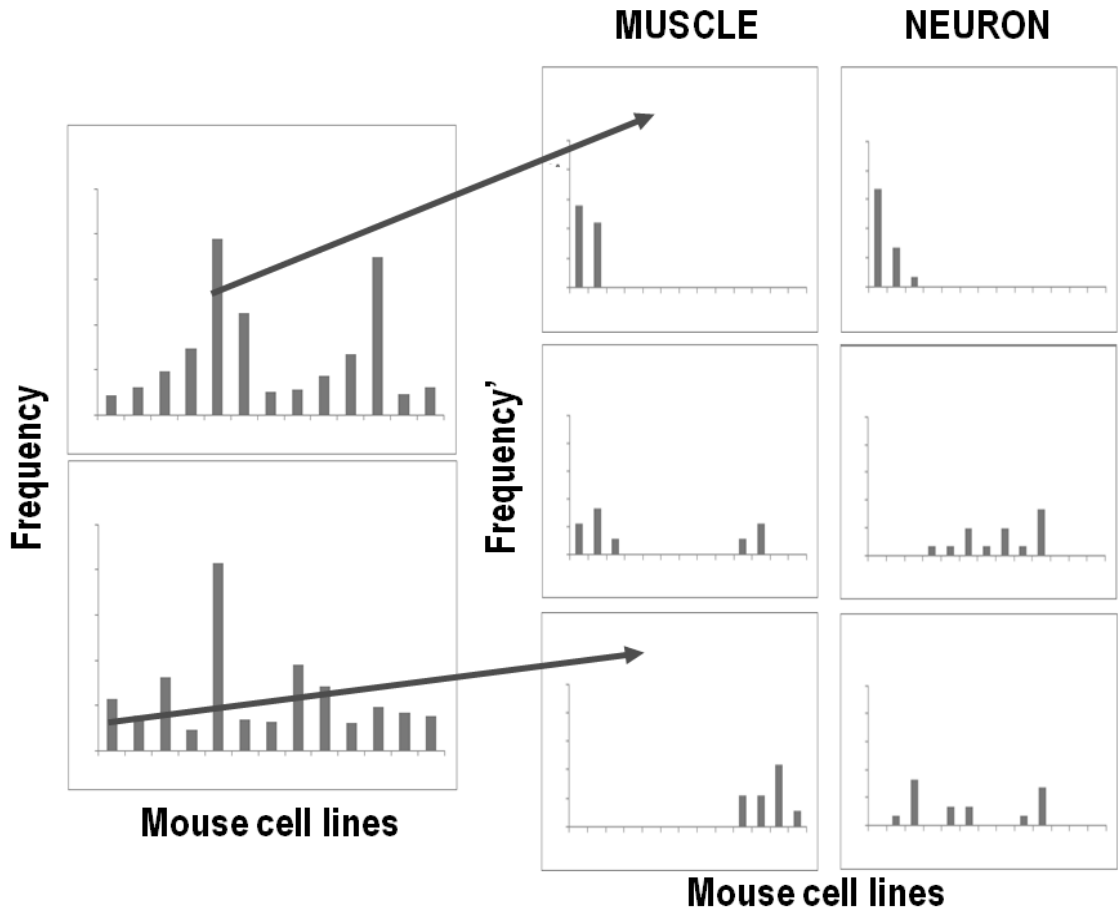

**Supplemental Figure 1.** Rank-order frequency method for characterizing reprogramming efficiency performance and repeatability between experiments. Graphs with frequency' represents the frequency of a given rank-order position across all replicates tested, while the graphs with the frequency y-axis represents frequency of each rank normalized by the inverse of each rank position and summed across all rank positions.

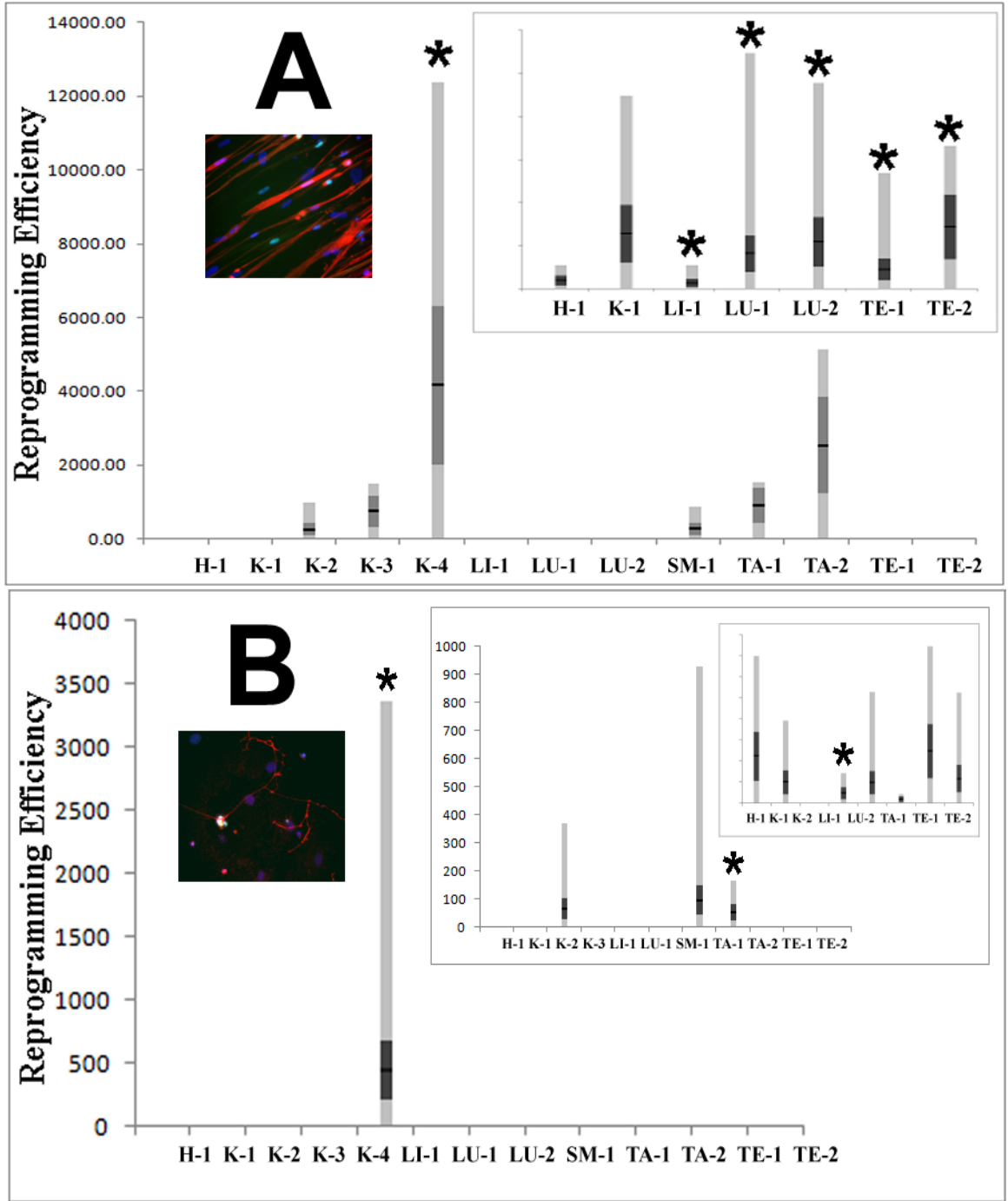

**Supplemental Figure 2.** Rate-based information extracted from reprogrammed cell lines. Results of a Poisson exact test for iSM mouse cells (A) and iN mouse cells (B).

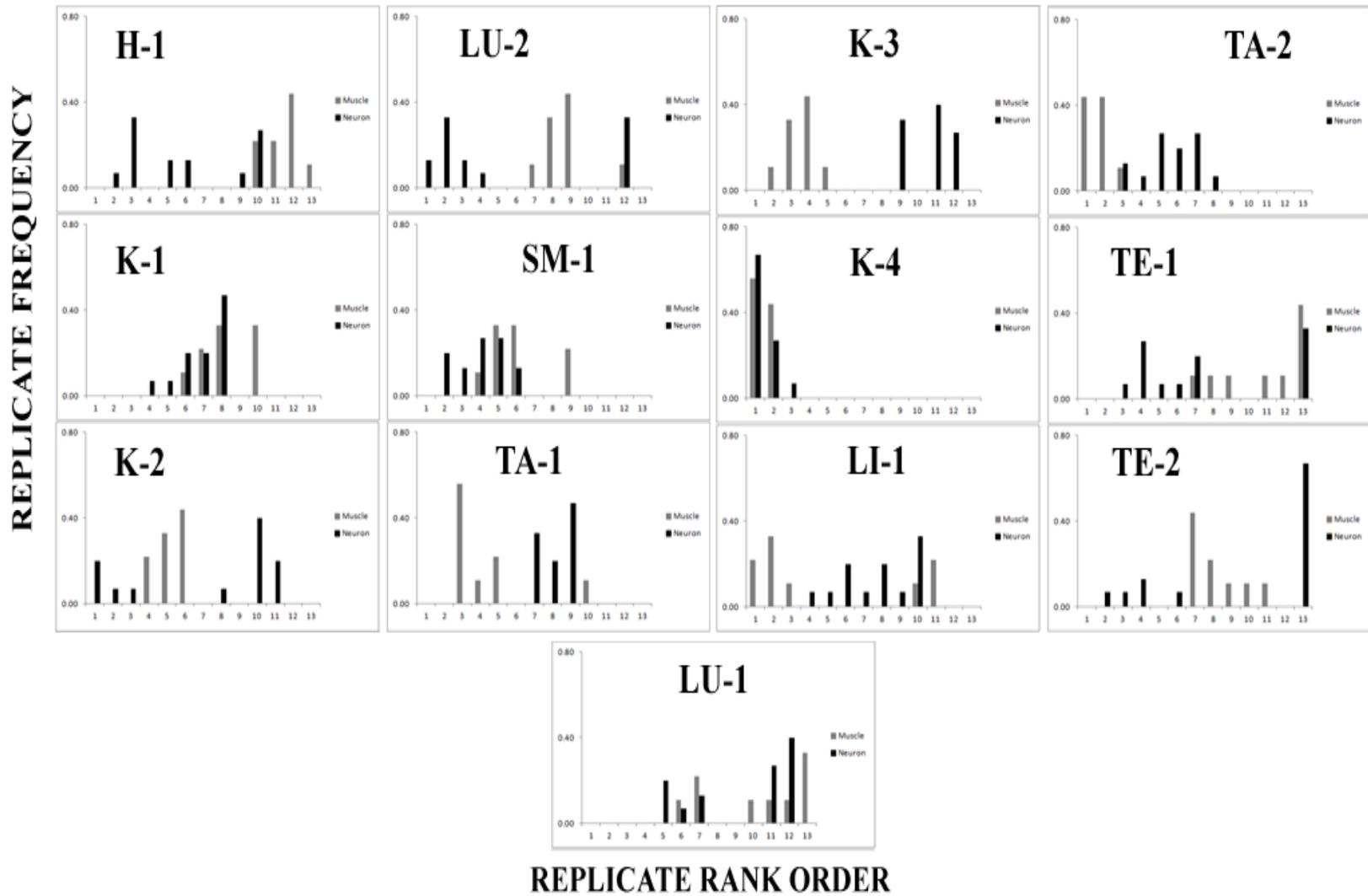

**Supplemental Figure 3.** Reprogramming bias as measured across 13 mouse cell lines (labeled) from different tissues and comparing iSM (gray bars) and iN (black bars) destination phenotypes.

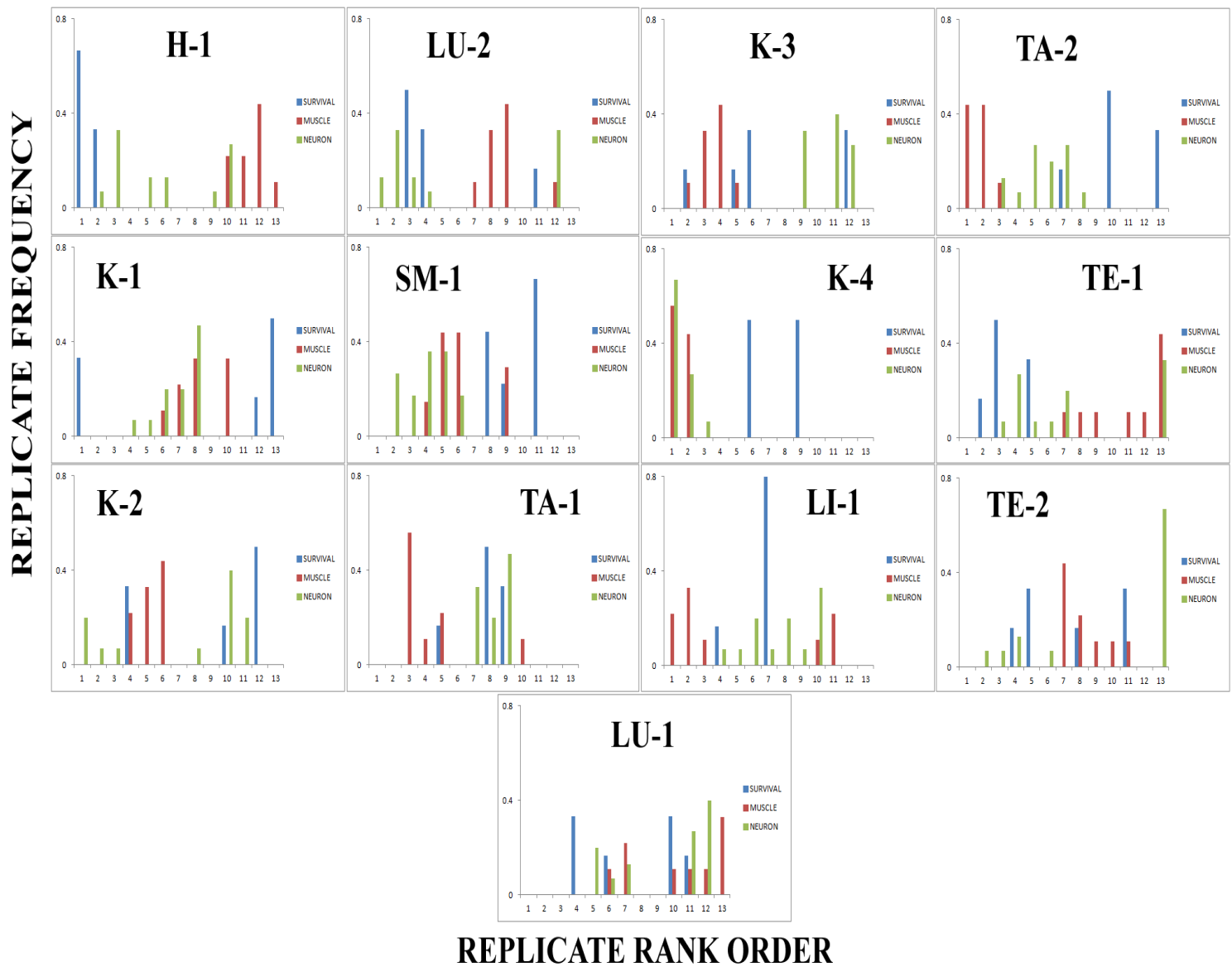

**Supplemental Figure 4.** Survivability analysis overlain with distribution of neurons and muscle reprogramming bias.